\documentclass[aps,prl,reprint,twocolumn,superscriptaddress,longbibliography,nofootinbib,secnumarabic]{revtex4-2}
\usepackage[T1]{fontenc}
\usepackage[utf8]{inputenc}
\usepackage[english]{babel}   
\usepackage{lmodern}
\usepackage{amsmath,amssymb,amsthm,bm,bbm}
\usepackage{siunitx}
\usepackage{xspace}
\usepackage{graphicx}
\usepackage{hyperref}
\usepackage{braket}
\usepackage{physics}        
\usepackage{enumitem}       
\usepackage{times}
\usepackage{siunitx}  
\usepackage{listings}
\usepackage{xcolor} 
\usepackage{mathtools }
\usepackage{tabularx}
\usepackage{tabularray}
\usepackage{threeparttable}
\usepackage{orcidlink} 
\usepackage{tcolorbox}
\usepackage{booktabs}

\newtcolorbox{draftrefs}{
  colback=gray!10,
  colframe=gray!50,
  title={References (for this section --- to verify later)},
  fonttitle=\bfseries
}

\begin{document}

\title{Basis- and Channel-Selective Quantum Photodetection}

\author{Mohamed Hatifi\,\orcidlink{0009-0005-3368-2751}}
\affiliation{Aix Marseille Univ, CNRS, Centrale M\'editerran\'ee, Institut Fresnel, Marseille, France}
\author{Brian Stout}
\affiliation{Aix Marseille Univ, CNRS, Institut Fresnel, Marseille, France}

\email{hatifi@fresnel.fr}

\begin{abstract}Photodetection converts optical quantum states into measurement events, but the usual electric-field response model becomes restrictive when the detector response is shaped by cavity, superconducting, or metamaterial engineering. We develop a generalized quantum photodetection framework in which electric and magnetic field amplitudes contribute coherently to the detection operator, and analyze it in a far-field two-source geometry, a two-mode single-photon setting, and a lossy resonant detector model. The far-field reference case exhibits complete detector-amplitude cancellation, absent in the electric-only Glauber response, while the single-photon model shows that the detector continuously rotates the effective measurement basis and controls the first-order visibility via an exact closed-form law. In the resonant realization, a monitored radiative output channel can be dark while the detector remains internally excited and absorptive, with unit absorption of the matched input mode at critical coupling. These results identify basis-selective readout and channel-selective absorption as experimentally relevant signatures of engineered electric--magnetic photodetection.
\end{abstract}

\maketitle

\section{Introduction}
\label{sec:introduction}

Photodetection is the operation through which optical quantum states become experimentally accessible. It converts field amplitudes into detector events and thereby determines which coherences, correlations, and mode occupations can be read out in quantum sensing, communication, and control. In standard quantum optics, this role is formalized by Glauber's theory of optical coherence, where the primary detection probability is expressed in terms of a normally ordered electric-field correlation function \cite{glauber1963,loudon2000,mandel1995,barnett2002}. This framework remains foundational because it captures a wide range of photonic measurements with a compact and predictive formalism. The question becomes less straightforward when the detector itself is an engineered electromagnetic object, as in superconducting nanowire detectors, cavity-assisted absorbers, and structured photonic or metamaterial platforms \cite{landy2008,hadfield2009,eisaman2011,natarajan2012,esmaeilzadeh2021}. How should photodetection be described when the measured response cannot be reduced, a priori, to a scalar electric sensitivity?

This question is motivated by the way modern photodetectors and absorbers are designed. Superconducting nanowire detectors and cavity-integrated single-photon architectures use geometry, impedance matching, and field confinement to shape how incoming radiation couples to dissipative degrees of freedom \cite{natarajan2012,holzman2019,steinhauer2021,venza2025}. Related work in structured light--matter environments further illustrates how engineered optical, microwave, and reservoir couplings can reshape fluorescence signatures, generated field states, and accessible quantum correlations \cite{hatifi2022b,kani2025b,hatifi2026,hatifi2026a}. Structured photonic and metamaterial platforms provide a complementary setting in which local electric and magnetic responses, field localization, and absorption can be engineered at subwavelength scales \cite{landy2008,chen2012,St15,Optinter2016,sureshkumar2021}. In such systems, the effective detector response depends on the local electromagnetic structure near the active region. Standard electric-field photodetection remains appropriate in many regimes, but these platforms make it timely to ask whether a coherently engineered electric--magnetic response can define a distinct quantum measurement. Several ingredients of this problem are already understood in classical scattering theory, where electric and magnetic dipolar responses, magnetoelectric point scatterers, and interference between electric and magnetic radiation channels provide standard tools for controlling scattering and absorption \cite{kerker1983,sersic2011,garcia-etxarri2011,geffrin2012}. Huygens-type and impedance-engineered structures extend this logic to designed photonic platforms, where balanced electric and magnetic responses are used to control reflection, transmission, and wavefront shaping \cite{pfeiffer2013,decker2015}. Related ideas have also been developed in quantum-optical settings, where collective atomic modes can synthesize effective magnetic responses, and Huygens surfaces \cite{paniagua-dominguez2016, ballantine2020}. These works establish electric--magnetic interference as a robust mechanism for controlling electromagnetic scattering. What remains to be formulated for the present purpose is a compact quantum photodetection model in which a coherent electric--magnetic detector response acts directly as a measurement operator, and in which radiative darkness and internal absorption can be treated within the same operational framework.
\\
\indent In this work, we introduce a generalized photodetection operator in which electric and magnetic field amplitudes contribute coherently to the detector excitation. The model is first applied to a two-dipole far-field reference geometry, which isolates the cancellation mechanism and shows how a complete null point can arise outside the electric-only Glauber model. We then analyze a two-mode single-photon field, where the same detector parameter acquires a direct measurement-theoretic meaning: it continuously rotates the effective readout basis and controls the detected first-order visibility through a closed-form law. Finally, we introduce a lossy resonant realization that separates external radiative outputs from internal dissipation, showing that a monitored output channel can be dark (carrying no outgoing radiation due to destructive interference) while the detector remains excited and absorptive, with unit absorption of the matched input mode at critical coupling. The resulting framework treats generalized photodetection as a design principle for engineered quantum detectors. A coherent electric--magnetic response can determine which single-photon superposition is measured and which radiative channel is suppressed or accessed. This provides concrete observables for future implementations: detector-controlled visibility, transition between interference-sensitive and path-sensitive readout, dark monitored output at balanced electric--magnetic coupling, and critical-coupling absorption. The theory, therefore, connects a foundational question in quantum measurement to experimentally motivated capabilities for basis-selective and channel-selective photodetection in engineered quantum platforms.

\section{Generalized photodetection and far-field reference case}
\label{sec:generalized_reference}

\subsection{Generalized electric--magnetic detection operator}
\label{subsec:formalism}

In the usual Glauber description, the lowest-order photodetection probability is proportional to the normally ordered electric-field correlation \cite{glauber1963,kelley1964,loudon2000,mandel1995,barnett2002}
\begin{equation}
P_{\mathrm{G}}(\mathbf{r},t)=
s^2
\left\langle
\widehat{\mathbf{E}}^{(-)}(\mathbf{r},t)\cdot
\widehat{\mathbf{E}}^{(+)}(\mathbf{r},t)
\right\rangle ,
\label{eq:GlauberRule}
\end{equation}
where \(s\) denotes an empirical detector sensitivity. Here \(\widehat{\mathbf{E}}^{(+)}\) denotes the positive-frequency part of the electric-field operator, containing the photon annihilation operators, while \(\widehat{\mathbf{E}}^{(-)}=[\widehat{\mathbf{E}}^{(+)}]^{\dagger}\) is its negative-frequency adjoint, containing the corresponding creation operators. This expression is appropriate when the detector response can be represented, to leading order, by an electric-dipole coupling to the positive-frequency electric field. The overall sensitivity \(s\) fixes the absolute count-rate scale; in what follows, we focus on the normalized detection patterns and on the field combinations sampled by the detector. To formulate an effective detector response that can also include magnetic sensitivity \cite{tanimura2014}, we first put the electric and magnetic field amplitudes in the same units by defining
\begin{equation}
\widehat{\mathbf{F}}^{(+)}(\mathbf{r},t)
\equiv
c\,\widehat{\mathbf{B}}^{(+)}(\mathbf{r},t),
\label{eq:Fdef_general}
\end{equation}
equivalently \(\widehat{\mathbf{F}}^{(+)}=Z_0\widehat{\mathbf{H}}^{(+)}\) in vacuum. The field \(\widehat{\mathbf{F}}^{(+)}\) has the same units as \(\widehat{\mathbf{E}}^{(+)}\), so that electric and magnetic contributions can be combined with a dimensionless relative response parameter. We consider a local detector operating in the linear-response regime, whose excitation amplitude is described by the positive-frequency operator
\begin{equation}
\widehat{\mathcal O}(\mathbf{r},t)=
\mathbf{u}_{e}^{*}\!\cdot\widehat{\mathbf{E}}^{(+)}(\mathbf{r},t)
+
\zeta\,
\mathbf{u}_{m}^{*}\!\cdot\widehat{\mathbf{F}}^{(+)}(\mathbf{r},t).
\label{eq:GenOp}
\end{equation}
Here \(\mathbf{u}_{e}\) and \(\mathbf{u}_{m}\) are unit vectors specifying the electric and magnetic polarization directions to which the detector is sensitive, while \(\zeta\in\mathbb{C}\) encodes the relative amplitude and phase of the magnetic response with respect to the electric one. The corresponding detection probability is then taken as
\begin{equation}
P(\mathbf{r},t)=
\left\langle
\widehat{\mathcal O}^{\dagger}(\mathbf{r},t)
\widehat{\mathcal O}(\mathbf{r},t)
\right\rangle .
\label{eq:GenProb}
\end{equation}
For \(\zeta=0\), Eq.~\eqref{eq:GenProb} reduces to the electric-only Glauber form in Eq.~\eqref{eq:GlauberRule}, up to the overall calibration factor. 

For \(\zeta\neq 0\), the detector samples a coherent electric--magnetic amplitude, and the probability contains interference terms between the two electromagnetic sectors. Equations~\eqref{eq:GenOp} and \eqref{eq:GenProb} should be understood as an effective photodetection model rather than as a device-specific microscopic Hamiltonian. They retain the operator structure of standard photodetection while allowing the detector to select a generalized electromagnetic amplitude. The rest of the paper examines the consequences of this single extension in three settings: a far-field reference geometry, a two-mode single-photon measurement, and a lossy resonant detector with distinct radiative and internal channels.
\begin{figure*}[t!]
\centering
\includegraphics[width=\textwidth]{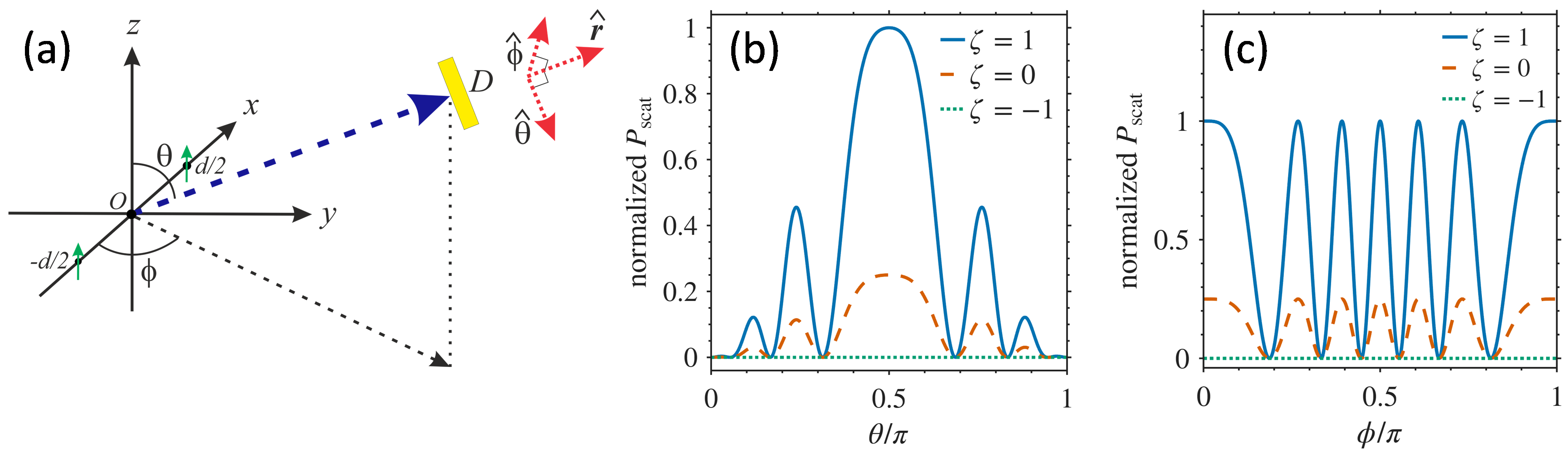}
\caption{
\textbf{Two-dipole far-field reference geometry for generalized electric--magnetic photodetection.}
(a) Two in-phase Hertzian dipoles with moments oriented along \(\hat{\mathbf z}\) are placed at positions \(\pm d/2\) and observed by a far-field detector \(D\). The local detector frame resolves the radiated electric and magnetic components through the generalized detection amplitude \(\widehat{\mathcal O}\propto \mathbf u_e\!\cdot\!\widehat{\mathbf E}^{(+)}+\zeta\,\mathbf u_m\!\cdot\!\widehat{\mathbf F}^{(+)}\), with \(\widehat{\mathbf F}^{(+)}=c\widehat{\mathbf B}^{(+)}\).
(b,c) Normalized radiative detection probability \(P_{\rm scat}\) for the polar cut \(\phi=0\) and the azimuthal cut \(\theta=\pi/2\), respectively, for \(d=3\lambda\). The electric-only Glauber response corresponds to \(\zeta=0\), while \(\zeta=1\) gives constructive electric--magnetic addition and \(\zeta=-1\) gives complete cancellation of the monitored radiative detection amplitude. Because the projected electric and magnetic far-field amplitudes are proportional in this geometry, the generalized detector changes the detection weight but not the fringe positions.
}
\label{fig:farfield_reference}
\end{figure*}
\subsection{Two-dipole far-field reference geometry}
\label{subsec:twodipole}

We first apply the generalized detector to a simple reference geometry in which the field interference is analytically transparent. Two in-phase Hertzian dipoles are placed at positions $\pm(d/2)\hat{\mathbf x}$, with identical dipole moments oriented along $\hat{\mathbf z}$. The detector is located in the far field along the direction specified by the spherical angles $(\theta,\phi)$, as sketched in Fig.~\ref{fig:farfield_reference}-(a). In the radiation zone, the total field is the coherent sum of the two dipole contributions \cite{jackson2009,novotny2012}. Omitting common radial, frequency, and normalization factors, the angular dependence of the electric field and of the rescaled magnetic field $\mathbf F=c\mathbf{B}$ is
\begin{align}
\mathbf{E}(\theta,\phi)
&\propto
\hat{\boldsymbol{\theta}}\,
\sin\theta\,
\cos\!\left[\frac{\delta(\theta,\phi)}{2}\right],
\label{eq:E_dip}
\\
\mathbf{F}(\theta,\phi)
&\propto
\hat{\boldsymbol{\phi}}\,
\sin\theta\,
\cos\!\left[\frac{\delta(\theta,\phi)}{2}\right],
\label{eq:F_dip}
\end{align}
with
\begin{equation}
\delta(\theta,\phi)=kd\sin\theta\cos\phi .
\label{eq:delta_dip}
\end{equation}
Here $\delta(\theta,\phi)$ is the far-field phase difference between the two source contributions. The factor $\cos[\delta(\theta,\phi)/2]$ is therefore the usual two-source interference amplitude. For a purely electric detector, Eq.~\eqref{eq:GlauberRule} gives
\begin{equation}
P_{\mathrm{G}}(\theta,\phi)\propto
\sin^2\theta\,
\cos^2\!\left[\frac{\delta(\theta,\phi)}{2}\right].
\label{eq:PG_dip}
\end{equation}
We now evaluate the generalized response using $\mathbf{u}_{e}=\hat{\boldsymbol{\theta}}$ and $\mathbf{u}_{m}=\hat{\boldsymbol{\phi}}$, which project onto the electric and magnetic far-field polarizations in Eqs.~\eqref{eq:E_dip} and \eqref{eq:F_dip}. Since the two projected amplitudes share the same angular factor, the generalized detection amplitude is proportional to
\begin{equation}
\sin\theta\,
\cos\!\left[\frac{\delta(\theta,\phi)}{2}\right](1+\zeta),
\end{equation}
and the corresponding radiative detection probability becomes
\begin{equation}
P_{\mathrm{scat}}(\theta,\phi)\propto
\sin^2\theta\,
\cos^2\!\left[\frac{\delta(\theta,\phi)}{2}\right]
\left(1+|\zeta|^2+2\Re[\zeta]\right).
\label{eq:Pscat}
\end{equation}
Equation~\eqref{eq:Pscat} gives the first useful lesson of the generalized detector model. The detector response can produce a complete amplitude cancellation at $\zeta=-1$, a possibility absent from the purely electric response in Eq.~\eqref{eq:PG_dip}, which only vanishes when $\delta=(2n+1)\pi$. At the same time, this reference geometry makes clear what is, and is not, being tested here. Because the projected electric and magnetic amplitudes are proportional at all observation angles, the generalized detector changes the overall detection weight but preserves the angular fringe structure inherited from the two sources. The far-field two-dipole example therefore isolates the electric--magnetic cancellation mechanism. To obtain a genuine detector-defined change of measurement basis, one must move to a geometry in which the electric and magnetic sectors address different mode superpositions, as done in the next section.

\section{Detector-defined single-photon interference}
\label{sec:quantum}

A detector-defined modification of first-order interference appears when the electric and magnetic sectors probe different superpositions of the same optical modes \cite{grangier1986,mandel1995}. The minimal setting is a pair of counterpropagating modes, as in a standing-wave or Fabry--P\'erot geometry. We therefore consider a one-dimensional field of fixed frequency $\omega=ck$, restricted to a right-moving mode $\hat a_R$ and a left-moving mode $\hat a_L$. At detector position $x$, the positive-frequency electric field and the rescaled magnetic field are written as
\begin{equation}
\widehat{\mathbf{E}}^{(+)}(x)=
\mathcal E
\left(
\hat a_R e^{ikx}+\hat a_L e^{-ikx}
\right)\hat{\mathbf y},
\label{eq:Eplus}
\end{equation}
\begin{equation}
\widehat{\mathbf{F}}^{(+)}(x)=
\mathcal E
\left(
\hat a_R e^{ikx}-\hat a_L e^{-ikx}
\right)\hat{\mathbf z},
\label{eq:Fplus}
\end{equation}
where $\mathcal E$ is the single-photon field amplitude. The relative minus sign in Eq.~\eqref{eq:Fplus} reflects the fact that the magnetic field changes sign when the propagation direction is reversed. As a result, the electric field probes the symmetric superposition of the two propagation directions, whereas the magnetic field probes the antisymmetric one. Choosing $\mathbf{u}_{e}=\hat{\mathbf y}$ and $\mathbf{u}_{m}=\hat{\mathbf z}$ in Eq.~\eqref{eq:GenOp}, the generalized detection operator becomes
\begin{equation}
\widehat{\mathcal O}_{\zeta}(x)=
\mathcal E
\left[
(1+\zeta)\hat a_R e^{ikx}
+
(1-\zeta)\hat a_L e^{-ikx}
\right].
\label{eq:Oexplicit}
\end{equation}
This expression makes explicit the detector's physical role: rather than simply rescaling a fixed response, $\zeta$ determines the specific superposition of the two propagation modes to which the detector is sensitive. We first evaluate this response for the single-photon state
\begin{equation}
\ket{\psi_{\phi}}=
\frac{1}{\sqrt2}
\left(
\hat a_R^{\dagger}+e^{i\phi}\hat a_L^{\dagger}
\right)\ket{0},
\label{eq:psi}
\end{equation}
where $\phi$ is the relative phase between the two components. The detection probability at position $x$ is
\begin{equation}
P_{\zeta}(x,\phi)=
\bra{\psi_{\phi}}
\widehat{\mathcal O}_{\zeta}^{\dagger}(x)
\widehat{\mathcal O}_{\zeta}(x)
\ket{\psi_{\phi}}.
\label{eq:Pdef}
\end{equation}
Using Eq.~\eqref{eq:Oexplicit}, the action of the detector on the single-photon state produces a vacuum amplitude,
\begin{equation}
\widehat{\mathcal O}_{\zeta}(x)\ket{\psi_{\phi}}
=
\mathcal A_{\zeta}(x,\phi)\ket{0},
\label{eq:Opsi_amp}
\end{equation}
where
\begin{equation}
\mathcal A_{\zeta}(x,\phi)
=
\frac{\mathcal E}{\sqrt2}
\left[
(1+\zeta)e^{ikx}
+
(1-\zeta)e^{-ikx}e^{i\phi}
\right].
\label{eq:Azeta}
\end{equation}
The detection probability is therefore simply
\begin{equation}
P_{\zeta}(x,\phi)
=
|\mathcal A_{\zeta}(x,\phi)|^2 .
\label{eq:Praw}
\end{equation}
The two terms in Eq.~\eqref{eq:Azeta} correspond to the right- and left-moving components as weighted by the detector. 

\noindent Their diagonal contributions fix the average count rate, while their cross-term carries the interference phase. Explicitly,
\begin{align}
P_{\zeta}(x,\phi)
&=
\frac{|\mathcal E|^2}{2}
\left[
|1+\zeta|^2
+
|1-\zeta|^2
\right.
\nonumber\\
&\hspace{1.0cm}
\left.
+
2\Re\!\left[
(1+\zeta)^{*}(1-\zeta)e^{-i(2kx-\phi)}
\right]
\right].
\label{eq:Pexpanded_step}
\end{align}
\begin{figure*}[t!]
\centering
\includegraphics[width=\textwidth]{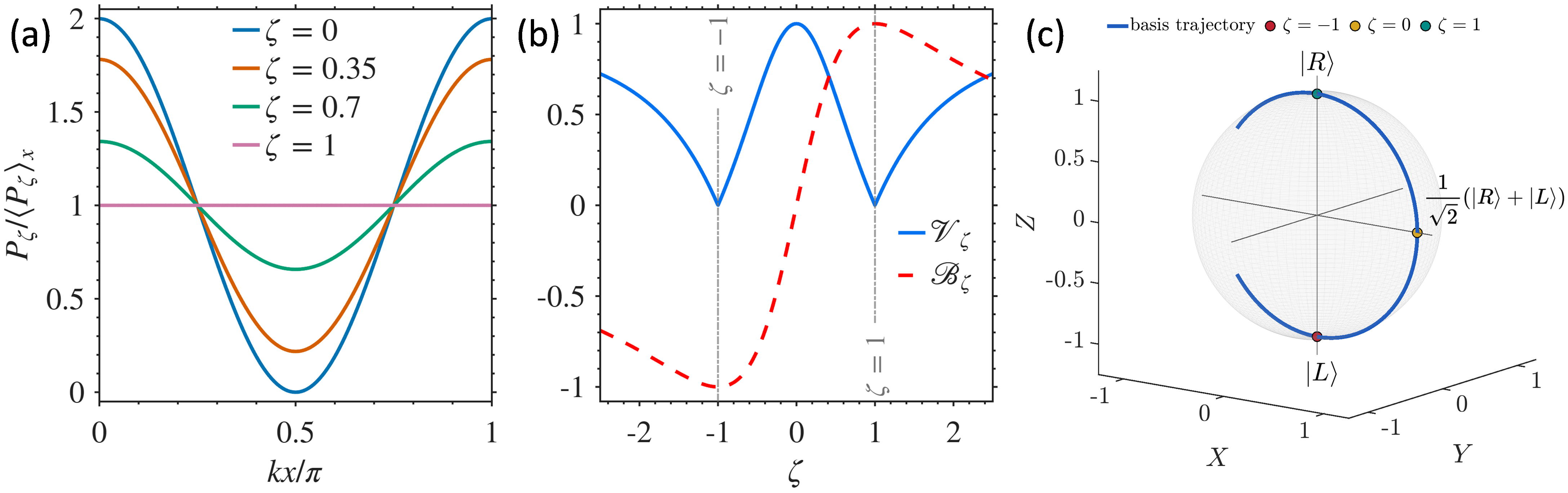}
\caption{
\textbf{Detector-defined single-photon interference in the two-mode geometry.}
(a) Normalized detection probability \(P_\zeta(x)/\langle P_\zeta\rangle_x\) for representative real values of the electric--magnetic response parameter \(\zeta\). As \(\zeta\) approaches the balanced value \(\zeta=1\), the detector becomes sensitive to a single propagation direction and the fringe visibility is suppressed, even though the single-photon state remains coherent.
(b) Visibility \(\mathcal V_\zeta\) and signed path bias \(\mathcal B_\zeta\) for real \(\zeta\). The vertical dashed lines mark \(\zeta=\pm1\), where the detector selects one of the two propagation directions and the visibility vanishes. The curves satisfy the complementarity relation \(\mathcal V_\zeta^2+\vert \mathcal B_\zeta\vert^2=1\), showing that the loss of visibility comes from a detector-controlled change of measurement basis rather than from decoherence.
(c) Bloch-sphere representation of the normalized detector-selected one-photon mode \(\ket{\widetilde{\chi}_\zeta}\). For real \(\zeta\), the measured mode moves along a meridian from \(\ket{L}\) at \(\zeta=-1\), through the balanced superposition \((\ket{R}+\ket{L})/\sqrt{2}\) at \(\zeta=0\), to \(\ket{R}\) at \(\zeta=1\). The detector parameter, therefore, continuously rotates the effective measurement basis in the one-photon Hilbert space.
}
\label{fig:detector_defined_interference}
\end{figure*}
Defining $Q_{\zeta}=(1+\zeta)^{*}(1-\zeta)=|Q_{\zeta}|e^{-i\delta_{\zeta}}$, the probability takes the standard fringe form
\begin{equation}
P_{\zeta}(x,\phi)=
|\mathcal E|^2\left(1+|\zeta|^2\right)
\left[
1+\mathcal V_{\zeta}
\cos\!\left(2kx-\phi-\delta_{\zeta}\right)
\right],
\label{eq:Pvis}
\end{equation}
with visibility
\begin{equation}
\mathcal V_{\zeta}
=
\frac{|Q_{\zeta}|}{1+|\zeta|^2}
=
\frac{|1-\zeta^2|}{1+|\zeta|^2}.
\label{eq:visibility}
\end{equation}
Equation~\eqref{eq:visibility} shows that the detected first-order interference depends jointly on the photon state and on the detector response encoded in $\zeta$. For $\zeta=0$, corresponding to a purely electric detector, one recovers the usual standing-wave pattern,
\begin{equation}
P_{0}(x,\phi)=
|\mathcal E|^2\left[1+\cos(2kx-\phi)\right],
\qquad
\mathcal V_{0}=1.
\label{eq:pureE}
\end{equation}
At the balanced points $\zeta=\pm1$, the detector becomes sensitive to a single propagation direction:
\begin{equation}
\widehat{\mathcal O}_{+1}(x)=2\mathcal E\,\hat a_R e^{ikx},
\qquad
\widehat{\mathcal O}_{-1}(x)=2\mathcal E\,\hat a_L e^{-ikx}.
\label{eq:balanced}
\end{equation}
The detection probability then becomes position independent,
\begin{equation}
P_{\pm1}(x,\phi)=2|\mathcal E|^2,
\qquad
\mathcal V_{\pm1}=0.
\label{eq:balancedP}
\end{equation}
The fringe suppression at \(\zeta=\pm1\) is therefore not a signature of decoherence. The photon remains in the pure coherent state \(\ket{\psi_\phi}\); the detector has instead selected an observable that is insensitive to the relative phase between the two propagation components. This point becomes especially transparent in the one-photon basis
\begin{equation}
\ket{R}=\hat a_R^{\dagger}\ket{0},
\qquad
\ket{L}=\hat a_L^{\dagger}\ket{0}.
\label{eq:RL}
\end{equation}
Within this one-photon subspace, the detector selects the normalized mode
\begin{equation}
\ket{\widetilde{\chi}_{\zeta}(x)}=
\frac{
(1+\zeta)e^{ikx}\ket{R}
+
(1-\zeta)e^{-ikx}\ket{L}
}{
\sqrt{2(1+|\zeta|^2)}
},
\label{eq:chi_norm}
\end{equation}
and the corresponding (Positive Operator-Valued Measure) POVM element reads
\begin{equation}
\Pi_{\zeta}^{(1)}(x)=
\widehat{\mathcal O}_{\zeta}^{\dagger}(x)\widehat{\mathcal O}_{\zeta}(x)
=
2|\mathcal E|^2(1+|\zeta|^2)\,
\ket{\widetilde{\chi}_{\zeta}(x)}\!\bra{\widetilde{\chi}_{\zeta}(x)},
\label{eq:POVM}
\end{equation}
where the superscript reminds us that the expression is restricted to the one-photon sector. 
\\ \\
In this form, the detector parameter $\zeta$ serves as a continuous control over the effective measurement basis, interpolating between a superposition-sensitive and a path-sensitive basis. The same interpretation extends immediately to a general one-photon qubit
\begin{equation}
\ket{\psi}=\alpha\ket{R}+\beta\ket{L},
\qquad
|\alpha|^2+|\beta|^2=1,
\label{eq:general_qubit}
\end{equation}
for which the detection probability becomes
\begin{equation}
P_{\zeta}(x)\propto
\left|
(1+\zeta)\alpha e^{ikx}
+
(1-\zeta)\beta e^{-ikx}
\right|^2.
\label{eq:general_prob}
\end{equation}
The generalized detector therefore implements a $\zeta$-dependent projective readout in the two-dimensional one-photon mode space. For real \(\zeta\), this interpolation can also be written as an exact complementarity relation \cite{wootters1979,jaeger1995,englert1996}. Defining the effective channel weights
\begin{equation}
w_R=(1+\zeta)^2,
\qquad
w_L=(1-\zeta)^2,
\label{eq:weights}
\end{equation}
we introduce the signed path bias
\begin{equation}
\mathcal B_{\zeta}
=
\frac{w_R-w_L}{w_R+w_L}
=
\frac{2\zeta}{1+\zeta^2}.
\label{eq:Bdef}
\end{equation}
The sign of \(\mathcal B_\zeta\) indicates which propagation direction is favored by the detector, while \(|\mathcal B_\zeta|\) gives the corresponding path distinguishability $\mathcal D_{\zeta}$. Together with Eq.~\eqref{eq:visibility}, this gives
\begin{equation}
\mathcal V_{\zeta}^{2}+\mathcal D_{\zeta}^{2}=1.
\label{eq:VD}
\end{equation}
The detector thus continuously transitions from interference-sensitive to path-sensitive measurement while the underlying photon state remains unchanged.

\section{Scattering-dark yet absorptive detector response}
\label{sec:dark_absorptive}

The previous section concerns how a generalized detector selects a measurement basis in the external optical field. A related, but physically distinct, question concerns energy flow: can destructive interference suppress a monitored radiative output channel while the detector still absorbs energy internally? Addressing this point requires separating two notions that are often conflated, namely reradiation into observable external channels and irreversible dissipation into internal detector degrees of freedom. The minimal model that captures this distinction is a single lossy resonance coupled radiatively through both electric and magnetic channels \cite{haus1984,gardiner1985c,wonjoosuh2004}. We therefore consider a localized resonant mode of frequency $\omega_0$, described in the weak-excitation regime by a complex amplitude $a(t)$. The mode couples radiatively to an electric channel and to a magnetic channel with rates $\gamma_e$ and $\gamma_m$, respectively, and decays irreversibly into internal detector degrees of freedom with rate $\gamma_i$; its total linewidth is thus given by $\Gamma=\gamma_e+\gamma_m+\gamma_i$. It is convenient to reorganize the two radiative channels into bright and dark superpositions,
\begin{equation}
\gamma_b=
\frac{\left(\sqrt{\gamma_e}+\sqrt{\gamma_m}\right)^2}{2},
\qquad
\gamma_d=
\frac{\left(\sqrt{\gamma_e}-\sqrt{\gamma_m}\right)^2}{2},
\label{eq:gamma_bd}
\end{equation}
so that $\gamma_b+\gamma_d=\gamma_e+\gamma_m\equiv\gamma_r$. The bright channel corresponds to the radiative superposition in which the electric and magnetic reradiation amplitudes add, whereas the dark channel corresponds to the superposition in which they subtract. The square roots in Eq.~\eqref{eq:gamma_bd} appear because input--output theory is written at the level of amplitudes \cite{gardiner1985c,wonjoosuh2004}: the radiative coupling amplitude is proportional to $\sqrt{\gamma}$, while the corresponding power rate is $\gamma$. We drive the detector with an incident mode matched to the bright radiative superposition. Its complex amplitude is denoted by \(s_{\rm in}\), with the input--output normalization chosen such that \(|s_{\rm in}|^2\) is the incident power flux. Writing the detuning as \(\Delta=\omega-\omega_0\), the resonant amplitude obeys
\begin{equation}
\dot a(t)=
\left(i\Delta-\frac{\Gamma}{2}\right)a(t)
+
\sqrt{\gamma_b}\,s_{\rm in}.
\label{eq:a_dot_dark}
\end{equation}
In the monochromatic steady state,
\begin{equation}
a=
\frac{\sqrt{\gamma_b}}{\Gamma/2-i\Delta}\,s_{\rm in}.
\label{eq:a_ss_dark}
\end{equation}
The outgoing amplitudes in the bright and dark external radiative channels are
\begin{equation}
s_b^{\rm out}=s_{\rm in}-\sqrt{\gamma_b}\,a,
\qquad
s_d^{\rm out}=-\sqrt{\gamma_d}\,a.
\label{eq:sout_bd}
\end{equation}
These relations have a simple interpretation. In the bright channel, the output is the coherent sum of the directly transmitted input amplitude and the field reradiated by the resonance. In the dark channel, there is no direct input contribution because the drive populates only the bright superposition; the entire output in that channel comes from the resonance's reradiation. By contrast, the internal decay rate $\gamma_i$ does not correspond to an external optical channel. Substituting Eq.~\eqref{eq:a_ss_dark} into Eq.~\eqref{eq:sout_bd} gives
\begin{equation}
s_b^{\rm out}=
\left(
1-\frac{\gamma_b}{\Gamma/2-i\Delta}
\right)s_{\rm in},
\label{eq:sb_out_explicit}
\end{equation}
\begin{equation}
s_d^{\rm out}=
-\frac{\sqrt{\gamma_b\gamma_d}}{\Gamma/2-i\Delta}\,s_{\rm in}.
\label{eq:sd_out_explicit}
\end{equation}
\begin{figure*}[t!]
\centering
\includegraphics[width=\textwidth]{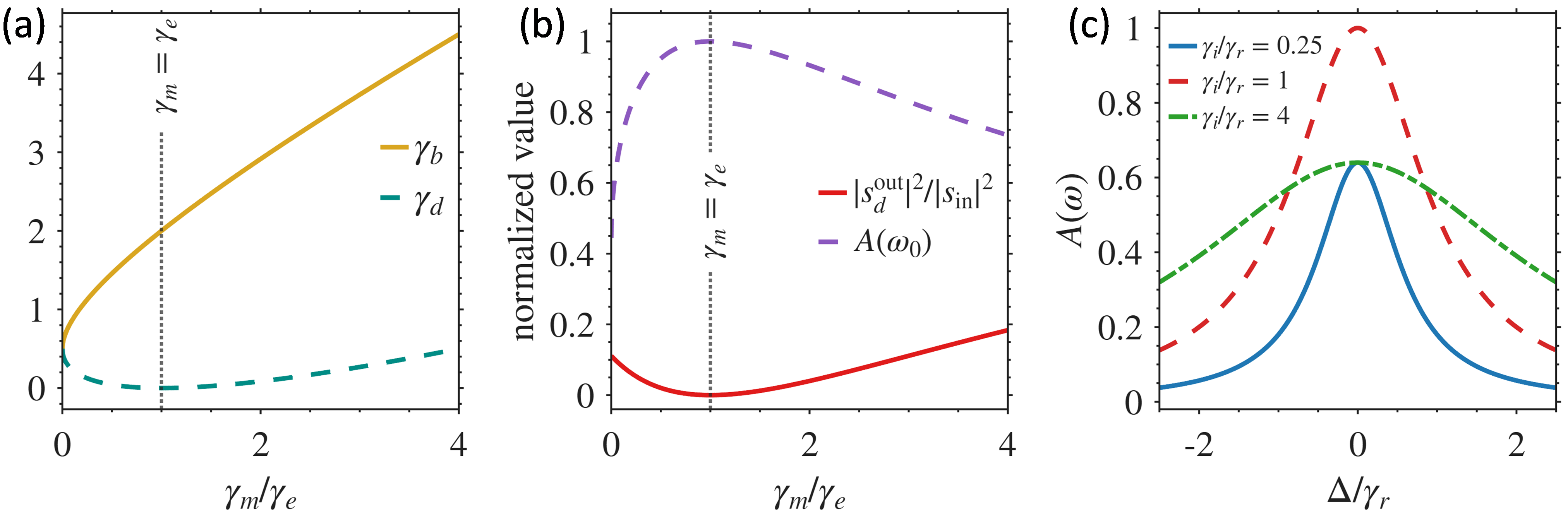}
\caption{
\textbf{Scattering-dark yet absorptive response of a lossy electric--magnetic detector resonance.}
(a) Bright and dark radiative rates \(\gamma_b\) and \(\gamma_d\) as functions of the magnetic-to-electric coupling ratio \(\gamma_m/\gamma_e\). At the balanced point \(\gamma_m=\gamma_e\), the dark-channel rate vanishes, \(\gamma_d=0\), while the bright channel remains coupled.
(b) Normalized dark-channel output \(|s_d^{\rm out}|^2/|s_{\rm in}|^2\) and resonant absorption \(A(\omega_0)\) for a matched bright input mode. At the balanced point, the monitored dark output is suppressed by destructive electric--magnetic reradiation, whereas the absorption remains finite and reaches unity here because the internal loss is chosen at critical coupling, \(\gamma_i=\gamma_r\).
(c) Absorption spectrum \(A(\omega)\) at balanced electric--magnetic coupling for three values of the internal-to-radiative loss ratio \(\gamma_i/\gamma_r\). The peak reaches unity at critical coupling, while in undercoupled and overcoupled regimes it shows lower peak absorption with different linewidths.
}
\label{fig:dark_absorptive}
\end{figure*}
Equation~\eqref{eq:sd_out_explicit} shows immediately that the dark radiative channel vanishes when
\begin{equation}
\gamma_d=0
\qquad \Longleftrightarrow \qquad
\gamma_e=\gamma_m.\nonumber
\label{eq:dark_condition}
\end{equation}
At this balanced electric--magnetic point, $s_d^{\rm out}=0$, while the resonance remains populated,
\begin{equation}
a=
\frac{\sqrt{\gamma_r}}{(\gamma_r+\gamma_i)/2-i\Delta}\,s_{\rm in}.
\label{eq:a_balanced}
\end{equation}
The disappearance of the monitored dark-channel output, therefore, does not signal the absence of interaction with the detector. It only signals destructive interference in one selected reradiated superposition. The absorbed fraction of the incident power follows from energy conservation in the input--output description \cite{gardiner1985c},
\begin{equation}
A(\omega)=
1-
\frac{|s_b^{\rm out}|^2+|s_d^{\rm out}|^2}{|s_{\rm in}|^2}.
\label{eq:Adef}
\end{equation}
A direct calculation yields
\begin{equation}
A(\omega)=
\frac{\gamma_b\gamma_i}{\Delta^2+(\Gamma/2)^2}.
\label{eq:A_general}
\end{equation}
The same result follows independently of the internal dissipative rate
\begin{equation}
R_{\rm abs}=\gamma_i|a|^2
=
\frac{\gamma_b\gamma_i}{\Delta^2+(\Gamma/2)^2}|s_{\rm in}|^2.
\label{eq:Rabs_general}
\end{equation}
Equation~\eqref{eq:Rabs_general} is the absorbed power inside the detector, whereas Eq.~\eqref{eq:A_general} is the absorbed fraction of the incoming power. Their agreement provides an explicit consistency check between the external input--output description and the internal dissipation dynamics. In the balanced regime, where $\gamma_d=0$ and $\gamma_b=\gamma_r$, Eqs.~\eqref{eq:A_general} and \eqref{eq:Rabs_general} reduce to
\begin{equation}
A(\omega)=
\frac{\gamma_r\gamma_i}
{\Delta^2+\left[(\gamma_r+\gamma_i)/2\right]^2},
\label{eq:A_balanced}
\end{equation}
and, at exact resonance, $A(\omega_0)=4\gamma_r\gamma_i/(\gamma_r+\gamma_i)^2$. The absorption is therefore strictly positive whenever both $\gamma_r$ and $\gamma_i$ are nonzero. At critical coupling $\gamma_i=\gamma_r$ one obtains $A(\omega_0)=1$ \cite{choi2001,chong2010,piper2014}. This result involves two distinct notions of darkness. The first is \emph{channel darkness}: at the balanced electric--magnetic point \(\gamma_e=\gamma_m\), the dark radiative output satisfies \(s_d^{\rm out}=0\). This cancellation only concerns one selected reradiated superposition. It does not imply that the detector is not excited, nor that all outgoing radiation has disappeared. In general, the detector can still reradiate through the bright output channel \(s_b^{\rm out}\). The second is \emph{absorption darkness}: at resonance and critical coupling, \(\Delta=0\) and \(\gamma_i=\gamma_r\), the remaining bright output also vanishes. Indeed, in the balanced regime one has
\begin{equation}
s_b^{\rm out}
=
\left(
1-\frac{\gamma_r}{(\gamma_r+\gamma_i)/2-i\Delta}
\right)s_{\rm in},
\end{equation}
so that \(s_b^{\rm out}=0\) when \(\Delta=0\) and \(\gamma_i=\gamma_r\). Since \(s_d^{\rm out}=0\) already holds at balance, no elastic radiative output remains and the matched incident power is dissipated internally. Thus, the detector can be dark in a monitored scattering channel while remaining internally populated and absorptive. The balanced electric--magnetic condition suppresses one selected reradiated channel through destructive interference between radiative amplitudes. Critical coupling suppresses the remaining bright output through destructive interference between the directly transmitted field and the field reradiated by the resonance. Absorption is then governed by the irreversible decay of the populated resonance into internal detector degrees of freedom at rate \(\gamma_i\). Radiative darkness and internal absorption are therefore not contradictory; they correspond to different observables of the same driven-dissipative detector.

\section{Platform relevance and implementation outlook}
\label{sec:implementation}

The theory developed above identifies two detector functionalities that are directly relevant to engineered quantum photodetection. The generalized operator \eqref{eq:GenOp} shows that a detector with coherent electric--magnetic response can select electromagnetic mode combinations beyond the purely electric response. The single-photon result of Sec.~\ref{sec:quantum} shows that this response can implement basis-selective readout in a two-mode one-photon Hilbert space. The resonant model of Sec.~\ref{sec:dark_absorptive} shows that the same principle can also produce channel-selective operation, where one monitored external radiative output is suppressed while internal absorption remains finite.

Several platform directions are compatible with this mechanism. In the optical regime, superconducting nanowire detectors, cavity-integrated absorbers, and structured photonic or metamaterial devices already rely on strong electromagnetic field shaping near the active region \cite{natarajan2012,akhlaghi2015,holzman2019,zhu2019,steinhauer2021,venza2025}. More generally, meta-atoms and impedance-engineered absorbers provide natural settings in which electric and magnetic couplings can be balanced or co-designed \cite{landy2008,chen2012,pfeiffer2013,decker2015,chen2012,chen2018a}. Microwave analog platforms are also attractive, since macroscopic structures with controlled electric and magnetic response can be engineered more explicitly, making the distinction between monitored radiative outputs and internal dissipation especially transparent \cite{landy2008,chen2018a,ballantine2020}. For a concrete structure, the parameter \(\zeta\) should be understood as an effective complex ratio between the magnetic and electric excitation amplitudes induced in the detector-active degree of freedom after projection onto the relevant local fields. In a resonant implementation, \(\gamma_e\) and \(\gamma_m\) would correspond to radiative linewidths associated with electric- and magnetic-like channels, while \(\gamma_i\) represents nonradiative or detector-internal decay. The predicted signatures, therefore, do not require direct access to microscopic detector amplitudes: they can be inferred from visibility measurements, elastic-output suppression, and absorption or count-rate spectra under controlled changes of geometry, impedance matching, or field localization.

The present work does not attempt a device-specific optimization. Instead, it identifies the effective parameters and observables that a concrete platform should control. The relevant detector parameter in the single-photon measurement problem is the complex electric--magnetic response ratio \(\zeta\), which determines the visibility \(\mathcal V_\zeta\) and the transition from interference-sensitive to path-sensitive readout. In the resonant problem, the relevant rates are the electric and magnetic radiative couplings \(\gamma_e\) and \(\gamma_m\), together with the internal loss rate \(\gamma_i\). The key experimental signatures are the suppression of a monitored dark output at the balanced point \(\gamma_e=\gamma_m\), finite internal absorption in that regime, and the critical-coupling absorption peak \(A(\omega_0)\) when \(\gamma_i=\gamma_r\) \cite{choi2001,chong2010,piper2014}. These quantities provide direct targets for future platform-specific modeling and measurements.

\section{Discussion}
\label{sec:discussion}

The results above give a unified operational picture of generalized photodetection. In the far-field two-dipole reference geometry, the detector response enters coherently at the level of the detection amplitude, resulting in a complete cancellation point. In the two-mode single-photon setting, the same response parameter acquires a direct quantum-measurement meaning: it rotates the effective measurement basis in the one-photon Hilbert space and controls the observed visibility. In the lossy resonant realization, the generalized response further determines which external radiative superpositions are bright or dark, while the internal dissipative channel remains available for absorption. This perspective shifts the role of the detector from a scalar efficiency factor to an active element of the measurement architecture. The detector determines which superpositions of modes are operationally accessible, which relative phases are visible, and which radiative channels are suppressed or monitored. The generalized electric--magnetic response can therefore be interpreted as a design principle for engineered quantum detectors: it provides a route to basis-selective readout in a single-photon mode space and channel-selective absorption in a driven-dissipative detector resonance.

The distinction between radiative output and internal absorption is especially important. A dark monitored channel does not necessarily indicate that the detector is inactive. In the resonant model, the incoming mode excites the detector, while destructive electric--magnetic interference suppresses one selected reradiated output. The stored excitation can still decay into the detector's internal degrees of freedom. This separates two notions that are often conflated: the darkness of an external scattering channel and the absence of light-matter interaction. The present work concerns the former, not a free-space dark fringe in which the incident field itself vanishes. The reduced models used here are chosen to isolate the measurement and energy-flow mechanisms in analytically transparent form. The far-field calculation serves as a reference case that identifies the electric--magnetic cancellation mechanism, while the two-mode calculation gives the smallest single-photon Hilbert space in which the detector changes the measured mode basis. Similarly, the resonant model captures the minimal separation between external reradiation and internal dissipation. A full multimode detector theory or a device-resolved microscopic model would be needed to extract platform-specific values of \(\zeta\), \(\gamma_e\), \(\gamma_m\), and \(\gamma_i\), but these extensions do not alter the mechanism established here.
\\
Within this scope, the theory identifies a clear set of experimentally relevant observables. The single-photon part predicts a detector-controlled visibility \(\mathcal V_\zeta\) and, for real \(\zeta\), an exact complementarity relation between visibility and detector-induced path bias \cite{wootters1979,jaeger1995,englert1996}. The resonant part predicts a balanced electric--magnetic point at which one monitored scattering channel is suppressed while internal absorption remains finite, with unit absorption of the matched input mode at critical coupling \cite{choi2001,chong2010,piper2014}. These quantities provide the natural targets for future platform-specific modeling and experimental tests of generalized quantum photodetection.

\section{Conclusion}
\label{sec:conclusion}

We have developed a generalized quantum photodetection framework in which a detector responds coherently to the amplitudes of electric and magnetic fields. The theory yields three connected results: a far-field cancellation mechanism beyond electric-only detection, a single-photon visibility law showing detector-controlled measurement-basis rotation, and a lossy resonant regime in which a monitored scattering channel is dark while the matched input mode can still be absorbed at critical coupling. These results identify coherent electric--magnetic response as a design principle for quantum detectors whose measured mode basis and accessible output channels are engineered rather than passively inherited. 

\noindent They motivate platform-specific studies in superconducting, nanophotonic, metamaterial, and microwave architectures, where the effective parameters \(\zeta\), \(\gamma_e\), \(\gamma_m\), and \(\gamma_i\) can be extracted and optimized for basis-selective and channel-selective photodetection.

\section*{Acknowledgments}
\noindent The authors thank Thomas Durt, Nicolas Bonod, Isam Ben Soltane, R\'emi Colom, Ross McPhedran, and Jacques Stout for stimulating discussions. This work was supported by the QuantAMU initiative at Aix-Marseille Universit\'e and the Institut Fresnel. M.H. acknowledges the support of the EU: the EIC Pathfinder Challenges 2022 call through the Research Grant 101115149 (project ARTEMIS)

\section*{Conflict of Interest}

The authors declare no conflict of interest.

%
%

\section*{Data Availability Statement}

No new experimental data were generated or analyzed in this study.


%

\end{document}